\documentstyle[12pt,epsf,epsfig]{article}
\def\beq{\begin{equation}}  
\def\eeq{\end{equation}}
\def\lsim{\mathrel{\rlap{\lower3pt\hbox{\hskip0pt$\sim$}}
    \raise1pt\hbox{$<$}}}         
\def\gsim{\mathrel{\rlap{\lower4pt\hbox{\hskip1pt$\sim$}}
    \raise1pt\hbox{$>$}}}         
\def\simlt{\mathrel{\raise.3ex\hbox{$<$\kern-.75em\lower1ex\hbox{$\sim$}}}}
\def\simgt{\mathrel{\raise.3ex\hbox{$>$\kern-.75em\lower1ex\hbox{$\sim$}}}}

\begin{document}
\begin{titlepage}

\begin{flushright}
hep-ph/0303249\\
TPI--MINN--03/09\\
UMN--TH--2134/03 \\
\end{flushright}
\begin{center}
\baselineskip25pt

\vspace{1cm}

{\Large\bf Higgs boson couplings to quarks with supersymmetric CP
and flavor violations}

\vspace{1cm}

{\sc
Durmu{\c s} A. Demir}
\vspace{0.3cm}

{\it Theoretical Physics Institute,
University of Minnesota, Minneapolis, MN 55455}

\end{center} 
\vspace{1cm} 
\begin{abstract} 
In minimal supersymmetric model (SUSY) with a light Higgs sector, 
explicit CP violation and most general flavor mixings
in the sfermion sector, integration of the superpartners out of the 
spectrum induces potentially large contributions to the Yukawa
couplings of light quarks via those of the heavier ones.  These 
corrections can be sizeable even for moderate values of 
$\tan\beta$, and remain nonvanishing even if all superpartners 
decouple. When the SUSY breaking scale is close to the electroweak scale, 
the Higgs exchange effects can compete with the gauge boson and 
box diagram contributions to rare processes, and their partial 
cancellations can lead to relaxation of the existing bounds on flavor 
violation sources. In this case there exist sizeable enhancements
in flavor--changing Higgs decays. When the superpartners completely
decouple, however, the Higgs mediation becomes the 
dominant SUSY contribution to rare processes the saturation 
of which, without a strong suppression of the flavor mixings,
prefers large $\tan\beta$ and certain ranges for the 
CP--odd phases. The decay rate of the lightest Higgs 
into light down quarks become comparable with that into the bottom quark. 
Moreover, the Higgs decay into the up quark is significantly enhanced.
There are observable implications for rare processes, atomic electric
dipole moments, and collider searches for Higgs bosons.

\end{abstract}

\end{titlepage}

The standard model of electroweak interactions (SM) has been
extremely successful in explaning all the available data. The
least understood aspects of the model concern the breaking
of {\it gauge}, {\it CP} and {\it flavor} symmetries. Indeed, the Higgs
boson mass and various parameters in the Yukawa matrices are
left to  experimental determination. Though the indications
at LEP for a light Higgs boson of mass  $\sim 115\ {\rm GeV}$ 
are encouraging a full construction of the symmetry--breaking 
sector, icluding possibly its CP properties, is to wait for the 
upgraded Tevatron or LHC. On the other hand, existing as well as
future data to come from the experiments on kaon, beauty and 
charmed hadrons will determine the structure of CP and flavor
violations.

The scalar sector, which is responsible for breaking the 
gauge symmetry, is quadratically sensitive to the UV cut--off and hence 
the model must be embedded into a UV--safe extension beyond
the ${\rm TeV}$ scale. Supersymmetry (SUSY) is the only 
weak--scale extension which stabilizes the Higgs sector 
against quadratic divergences and unifies the gauge couplings 
at high energies in agreement with the electroweak precision data. 
Quite generically, the SUSY models bring about novel sources
for CP and flavor violations through the soft breaking masses.
The main reason for SUSY flavor violation is that the fermions 
and sfermions are misaligned in the flavor space, and even if the flavor 
violation in the fermion sector is reduced to that of 
the CKM matrix the sfermion sector maintains its non--CKM structure.

The LR and RL=LR$^{\dagger}$ blocks of the sfermion mass--squared matrices are 
generated after the electroweak breaking with the maximal size ${\cal{O}}(m_t 
M_{SUSY})$. The nontrivial flavor structures of these blocks are dictated by
the Yukawa couplings ${\bf Y}_{u,d}$ and by the trilinear coupling 
matrices ${\bf Y}_{u,d}^{A}$ with
\begin{eqnarray}
\label{aterms}
\left({\bf Y}_{u}^{A}\right)_{i j} = \left({\bf Y}_{u}\right)_{i j} 
\left(A_{u}\right)_{i j} \:\:\: \mbox{and} \:\:\:  
\left({\bf Y}_{d}^{A}\right)_{i j} = \left({\bf Y}_{d}\right)_{i j} 
\left(A_{d}\right)_{i j}
\end{eqnarray} 
where $A_{u,d}$ are not necessarily unitary so that even their
diagonal entries contribute to CP--violating observables.
The LL and RR blocks 
are insensitive to electroweak breaking, and their texture is determined 
by the SUSY breaking pattern. In minimal SUGRA and its nonuniversal variants with CP violation, 
for instance, size and structure of flavor and CP violation 
in LL and RR blocks are dictated by the CKM matrix \cite{biz}. 
On the other hand, in SUSY GUTs with Yukawa unification
$e.g.$ SO(10), implementation of the see--saw mechanism
for neutrino masses implies sizeable flavor violation in the RR block, 
given the large mixings observed in  atmospheric neutrino 
data \cite{masiero}. Independent of specific realizations, 
the squark mass--squared matrices can be paramterized as 
\begin{eqnarray}
\label{massd}
\left(M_{D}^{2}\right)_{LL}=\left(\begin{array}{ccc}
M^{2}_{\tilde{d}_L} & M^{2}_{\tilde{d}_L \tilde{s}_L} &  M^{2}_{\tilde{d}_L 
\tilde{b}_L} \\\\
 M^{2}_{\tilde{s}_L \tilde{d}_L}&M^{2}_{\tilde{s}_L} & M^{2}_{\tilde{s}_L 
\tilde{b}_L}\\\\
M^{2}_{\tilde{b}_L \tilde{d}_L}&M^{2}_{\tilde{b}_L \tilde{s}_L} & 
M^{2}_{\tilde{b}_L} 
\end{array}\right)
\:\:,\:\:\:\:\:\: 
\left(M_{D}^{2}\right)_{RR}=\left(\begin{array}{ccc}
M^{2}_{\tilde{d}_R} & M^{2}_{\tilde{d}_R \tilde{s}_R} &  M^{2}_{\tilde{d}_R
\tilde{b}_R} \\\\
 M^{2}_{\tilde{s}_R \tilde{d}_R}&M^{2}_{\tilde{s}_R} & M^{2}_{\tilde{s}_R
\tilde{b}_R}\\\\
M^{2}_{\tilde{b}_R \tilde{d}_R}&M^{2}_{\tilde{b}_R \tilde{s}_R} &
M^{2}_{\tilde{b}_R}
\end{array}\right)
\end{eqnarray}
in the bases $\{\tilde{d}_L,\tilde{s}_L, \tilde{b}_L\}$ and 
$\{\tilde{d}_R, \tilde{s}_R, \tilde{b}_R\}$, respectively. 
The same structure repeats for the up sector. The hermiticity
of the mass matrices, $\left(M_{D}^{2}\right)_{LL,RR}= 
\left(M_{D}^{2}\right)_{LL,RR}^{\dagger}$, allows CP
violation only in the off--diagonal entries.

In comparison to the SM amplitudes, the virtual effects of 
sparticles on the rare processes scale as $M_W/M_{SUSY}$ to 
appropriate power due to either their derivative coupling to
the vector bosons or the sensitivity of the particular amplitude to the electroweak 
breaking \cite{fcnc1,fcnc1p}. Similarly, the hadronic and
leptonic dipole moments scale as (fermion mass)/$M_{SUSY}^{2}$.
In this sense, various bounds on SUSY flavor and CP violation sources
from the current experimental data depend on how close $M_{SUSY}$
is the electroweak scale. Looking from a different 
channel, the FCNC couplings of $Z$ boson to fermions scale as 
$M_Z^{2}/M_{SUSY}^{2}$ for $Z$ boson decays \cite{zbs}, and SUSY effects 
become transparent only at collider energies $E\sim M_{SUSY}$. This
decoupling property of the SUSY effects does not hold for interactions
of Higgs bosons with fermions as their couplings to sfermions are
dictated by the soft--breaking sector. Consequently, gauge and Higgs 
bosons, considering their decays and productions as well as the 
FCNC processes they mediate, possess essential differences 
concerning their sensitivity to the SUSY breaking scale. Indeed, 
the contributions of the sparticles, even if they are too heavy to be produced 
directly at near--future colliders, to gauge (Higgs) boson
couplings to fermions are (are not) suppressed by $1/M_{SUSY}$.
This nondecoupling property of the Higgs bosons persists
unless the Higgs sector itself enters the decoupling regime
in which case the SM results are recovered \cite{hff}. Therefore,
when the SUSY Higgs sector is stabilized at the weak scale the
Higgs boson interactions with the standard matter provides
a direct access to SUSY even if it can be in the decoupling regime. The 
Higgs--mediated FCNC becomes sizeable when the vacuum expectation
values (VEV) of the Higgs fields are hierarchically split. This
regime of the parameter space is motivated by LEP constraints on the SUSY parameter
space and by the Yukawa--unified models like SO(10) \cite{so10,higgsfcnc,higgsfcnc1}.
Depending on what sparticles are contained in the light spectrum the weak--scale 
effective theory can vary from a two--doublet model to a full
SUSY model as two extremes. The EDM and FCNC constraints on
Higgs mediation can be strong for the former \cite{lp,higgsfcnc,higgsfcnc1}
whereas they can be milder for the latter \cite{pilaftsisx,bsgam}.

The purpose of this work is to compute the couplings of
Higgs bosons to quarks in the presence of SUSY CP and flavor
violation effects within the minimal SUSY model. It will be 
shown that there are parametrically
sizeable corrections to light quark Yukawas which imply novel properties: 
($i$) the present constraints from 
non--Higgs contributions to FCNC processes \cite{fcnc1,fcnc1p} can
be modified, ($ii$) the EDMs can probe CP violation from both flavor--blind 
and flavor--sensitive SUSY phases, ($iii$) the flavor--violating 
decay rates of the Higgs bosons can be comparable with the 
flavor--conserving ones, and ($iv$) the Higgs bosons can turn 
out to be totally blind to all quarks but the charm and the top.
These phenomena have observable signatures for experiments
at meson factories as well as Higgs searches at colliders.

In general, in models with two or more Higgs doublets suppression  
of the tree level FCNC is accomplished by imposing certain symmetries
In minimal SUSY, it is a U(1) symmetry under which all fields are
neutral except for ${\bf d}_{R}$ and $H_d$ which have identical charges.
This implies that the Higgs doublet $H_u$ ($H_d$) couples only to up (down) 
type quarks . However, the symmetry under concern  is broken at the loop level 
due to the soft SUSY--breaking masses \cite{so10,higgsfcnc}. Thus, the 
effective lagrangian describing the Higgs--quark interactions below $M_{SUSY}$
may be written as 
\begin{eqnarray}
\label{efflag}
-{\cal{L}}&=&\overline{{\bf d}_{R}}\ \left[ {\bf Y}_{d} - \gamma^d 
\right]\ H_{d}^{0}\ {\bf d}_{L}
          + \overline{{\bf d}_{R}}\ \Gamma^d\ H_{u}^{0\, \star}\ 
{\bf d}_{L}\nonumber\\
         &+& \overline{{\bf u}_{R}}\ \left[ {\bf Y}_{u} + \gamma^u 
\right]\ H_{u}^{0}\ {\bf u}_{L}
          - \overline{{\bf u}_{R}}\ \Gamma^u\ H_{d}^{0\, \star}\ 
{\bf u}_{L}\:\:+\:\:\: \mbox{h.c.}
\end{eqnarray}
where, at tree level, flavor and CP violations are entirely 
determined by the Yukawa matrices ${\bf Y}_{d}$ and 
${\bf Y}_{u}$ whose simultaneous digonalization leads to the CKM matrix as the 
only observable effect. Therefore, without loss of generality, one can 
choose an appropriate basis for ${\bf Y}_{d,u}$ such as the down quark diagonal one
\begin{eqnarray}
\label{yukawa}
{\bf Y}_{d}= \left(\begin{array}{ccc}
 h_d&0&0\\\\
0&h_s&0\\\\
0&0&h_b\end{array}\right)\:\:\:\: ,\:\:\: {\bf Y}_{u}=\left(\begin{array}{ccc}
h_u&0&0\\\\
0&h_c&0\\\\
0&0&h_t\end{array}\right)\cdot V^0
\end{eqnarray} 
where $h_i$ and $V^0$ are tree level Yukawa couplings and the CKM matrix, respectively. 

The nonholomorphic Yukawa structures $\gamma^{u,d}$ and $\Gamma^{u,d}$ in (\ref{efflag})
result from integrating out the heavy degrees of freedom which may include the entire sparticle
spectrum or part of it. The dominant contributions to these SUSY threshold effects can be 
gathered by employing the SU(2)$_L\times$U(1)$_Y$ symmetric limit and neglecting their 
gauge couplings ($c.f.$ \cite{higgsfcnc1} for a discussion the electroweak breaking effects). 
Then the electroweak breaking occurs after integrating out 
the sparticles. In this limit, the LR and RL blocks of the sfermion mass matrices 
vanish so do the self--energy corrections on the quark lines. Hence, $\gamma^{u,d}$
and $\Gamma^{u,d}$ are generated by the vertex diagrams meditated by
gluino--squark and Higgsino--squark loops. Using the Yukawa bases
(\ref{yukawa}) in the trilinear couplings (\ref{aterms}) and
relabelling the quarks and squarks as $\{d,s,b\}\equiv \{d^1,d^2,d^3\}$
and $\{u,c,t\}\equiv \{u^1,u^2,u^3\}$, one finds 
\begin{eqnarray}
\label{diagd}
\gamma^{d}_{i i} &=& \frac{2 \alpha_s}{3\pi}\ 
\left({\bf Y}_{d}^{A}\right)_{i i} M_{g}^{\star}\ 
I_3\left(M_{\tilde{d}^{i}_L}^{2}, 
M_{\tilde{d}^{i}_R}^{2}, |M_{g}|^{2}\right)\nonumber\\
&+&\frac{2 \alpha_s}{3\pi} \sum_{j=1}^{3}
\left({\bf Y}_{d}^{A}\right)_{j j} M_{g}^{\star}\ M_{\widetilde{D}}^{4}
I_5\left(M_{\tilde{d}^{i}_L}^{2},M_{\tilde{d}^{j}_L}^{2},
M_{\tilde{d}^{j}_R}^{2}, M_{\tilde{d}^{i}_R}^{2}, |M_{g}|^{2}\right)\
\left(\delta^{d}_{ij}\right)_{RR} \left(\delta^{d}_{ji}\right)_{LL}\nonumber\\
&+&\frac{\left({\bf Y}_{d}\right)_{i i}}{(4\pi)^{2}}\ \sum_{j=1}^{3}\ 
\left({\bf Y}_{u}^{\dagger}\right)_{i j} 
\left({\bf Y}_{u}\right)_{j i} |\mu|^{2}\ I_3\left(M_{\tilde{u}^{j}_R}^{2}, 
M_{\tilde{u}^{i}_L}^{2}, |\mu|^{2}\right)\nonumber\\
\Gamma^{d}_{i i} &=& \frac{2 \alpha_s}{3\pi}\ \left({\bf Y}_{d}\right)_{i i}
\mu^{\star} M_{g}^{\star}\ I_3\left(M_{\tilde{d}^{i}_L}^{2},
M_{\tilde{d}^{i}_R}^{2}, |M_{g}|^{2}\right)\nonumber\\
&+&\frac{2 \alpha_s}{3\pi} \sum_{j=1}^{3}
\left({\bf Y}_{d}\right)_{j j} \mu^{\star} M_{g}^{\star}\ M_{\widetilde{D}}^{4}
I_5\left(M_{\tilde{d}^{i}_L}^{2},M_{\tilde{d}^{j}_L}^{2},
M_{\tilde{d}^{j}_R}^{2}, M_{\tilde{d}^{i}_R}^{2}, |M_{g}|^{2}\right)\
\left(\delta^{d}_{ij}\right)_{RR} \left(\delta^{d}_{ji}\right)_{LL}\nonumber\\
&+&\frac{\left({\bf Y}_{d}\right)_{i i}}{(4\pi)^{2}}\ \sum_{j=1}^{3}\ 
\left({\bf Y}_{u}^{A\, \dagger}\right)_{i j}
\left({\bf Y}_{u}\right)_{j i} \mu^{\star}\ I_3\left(M_{\tilde{u}^{j}_R}^{2},
M_{\tilde{u}^{i}_L}^{2}, |\mu|^{2}\right)
\end{eqnarray}
for the diagonal elements, and
\begin{eqnarray}
\label{offdiagd}
\gamma^{d}_{i j} &=& \frac{2 \alpha_s}{3\pi}\
\left({\bf Y}_{d}^{A}\right)_{i i} M_{g}^{\star} M_{\widetilde{D}}^{2}\
I_4\left(M_{\tilde{d}^{j}_L}^{2},
M_{\tilde{d}^{i}_L}^{2}, M_{\tilde{d}^{i}_R}^{2}, |M_{g}|^{2}\right)\: 
\left(\delta^{d}_{ij}\right)_{LL}\nonumber\\
&+&\frac{2 \alpha_s}{3\pi}\
\left({\bf Y}_{d}^{A}\right)_{j j} M_{g}^{\star} M_{\widetilde{D}}^{2}\
I_4\left(M_{\tilde{d}^{j}_L}^{2},
M_{\tilde{d}^{j}_R}^{2}, M_{\tilde{d}^{i}_R}^{2}, |M_{g}|^{2}\right)\:
\left(\delta^{d}_{ij}\right)_{RR}\nonumber\\
&+&\frac{\left({\bf Y}_{d}\right)_{i i}}{(4\pi)^{2}}\ 
\left({\bf Y}_{u}^{\dagger}\right)_{i j}
\left({\bf Y}_{u}\right)_{j j} |\mu|^{2}\ I_3\left(M_{\tilde{u}^{j}_R}^{2},
M_{\tilde{u}^{i}_L}^{2}, |\mu|^{2}\right)\nonumber\\
&+&\frac{\left({\bf Y}_{d}\right)_{i i}}{(4\pi)^{2}}\
\left({\bf Y}_{u}^{\dagger}\right)_{i i}
\left({\bf Y}_{u}\right)_{i j} |\mu|^{2}\ I_3\left(M_{\tilde{u}^{i}_R}^{2},
M_{\tilde{u}^{i}_L}^{2}, |\mu|^{2}\right)\nonumber\\
&+&\frac{\left({\bf Y}_{d}\right)_{i i}}{(4\pi)^{2}}\
\left({\bf Y}_{u}^{\dagger}\right)_{j j}
\left({\bf Y}_{u}\right)_{j j} |\mu|^{2} M_{\widetilde{U}}^2\
I_4\left(M_{\tilde{u}^{j}_R}^{2}, M_{\tilde{u}^{j}_L}^{2}, 
M_{\tilde{u}^{i}_L}^{2}, |\mu|^{2}\right)\: 
\left(\delta^{u}_{ij}\right)_{LL}\nonumber\\
&+&\frac{\left({\bf Y}_{d}\right)_{i i}}{(4\pi)^{2}}\
\left({\bf Y}_{u}^{\dagger}\right)_{i i}
\left({\bf Y}_{u}\right)_{j j} |\mu|^{2} M_{\widetilde{U}}^2\
I_4\left(M_{\tilde{u}^{j}_R}^{2}, M_{\tilde{u}^{i}_R}^{2},
M_{\tilde{u}^{i}_L}^{2}, |\mu|^{2}\right)\:
\left(\delta^{u}_{ij}\right)_{RR}\nonumber\\
\Gamma^{d}_{i j} &=& \frac{2 \alpha_s}{3\pi}\
\left({\bf Y}_{d}\right)_{i i} \mu^{\star} M_{g}^{\star} M_{\widetilde{D}}^{2}\
I_4\left(M_{\tilde{d}^{j}_L}^{2},
M_{\tilde{d}^{i}_L}^{2}, M_{\tilde{d}^{i}_R}^{2}, |M_{g}|^{2}\right)\: 
\left(\delta^{d}_{ij}\right)_{LL}\nonumber\\
&+&\frac{2 \alpha_s}{3\pi}\
\left({\bf Y}_{d}\right)_{j j} \mu^{\star} M_{g}^{\star} M_{\widetilde{D}}^{2}\
I_4\left(M_{\tilde{d}^{j}_L}^{2},
M_{\tilde{d}^{j}_R}^{2}, M_{\tilde{d}^{i}_R}^{2}, |M_{g}|^{2}\right)\:
\left(\delta^{d}_{ij}\right)_{RR}\nonumber\\
&+&\frac{\left({\bf Y}_{d}\right)_{i i}}{(4\pi)^{2}}\ 
\left({\bf Y}_{u}^{A\, \dagger}\right)_{i j}
\left({\bf Y}_{u}\right)_{j j} \mu^{\star}\ I_3\left(M_{\tilde{u}^{j}_R}^{2},
M_{\tilde{u}^{i}_L}^{2}, |\mu|^{2}\right)\nonumber\\
&+&\frac{\left({\bf Y}_{d}\right)_{i i}}{(4\pi)^{2}}\
\left({\bf Y}_{u}^{A\, \dagger}\right)_{i i}
\left({\bf Y}_{u}\right)_{i j} \mu^{\star}\ I_3\left(M_{\tilde{u}^{i}_R}^{2},
M_{\tilde{u}^{i}_L}^{2}, |\mu|^{2}\right)\nonumber\\
&+&\frac{\left({\bf Y}_{d}\right)_{i i}}{(4\pi)^{2}}\
\left({\bf Y}_{u}^{A\, \dagger}\right)_{j j}
\left({\bf Y}_{u}\right)_{j j}  \mu^{\star} M_{\widetilde{U}}^2\
I_4\left(M_{\tilde{u}^{j}_R}^{2}, M_{\tilde{u}^{j}_L}^{2}, 
M_{\tilde{u}^{i}_L}^{2}, |\mu|^{2}\right)\: 
\left(\delta^{u}_{ij}\right)_{LL}\nonumber\\
&+&\frac{\left({\bf Y}_{d}\right)_{i i}}{(4\pi)^{2}}\
\left({\bf Y}_{u}^{A\, \dagger}\right)_{i i}
\left({\bf Y}_{u}\right)_{j j} \mu^{\star} M_{\widetilde{U}}^2\
I_4\left(M_{\tilde{u}^{j}_R}^{2}, M_{\tilde{u}^{i}_R}^{2},
M_{\tilde{u}^{i}_L}^{2}, |\mu|^{2}\right)\:
\left(\delta^{u}_{ij}\right)_{RR}
\end{eqnarray}
for the off--diagonal elements. These expressions (\ref{diagd}) and (\ref{offdiagd}), 
with $i,j=1,2,3$, complete the radiative corrections to down quark interactions with Higgs fields.
Repeating a similar analysis for the up quark sector, one finds
\begin{eqnarray}
\gamma^{u}_{i i} &=& \frac{2 \alpha_s}{3\pi}\ 
\left({\bf Y}_{u}^{A}\right)_{i i} M_{g}^{\star}\ 
I_3\left(M_{\tilde{u}^{i}_L}^{2}, 
M_{\tilde{u}^{i}_R}^{2}, |M_{g}|^{2}\right)\nonumber\\
&+&\frac{2 \alpha_s}{3\pi} \sum_{j=1}^{3}
\left({\bf Y}_{u}^{A}\right)_{j j} M_{g}^{\star}\ M_{\widetilde{U}}^{4}
I_5\left(M_{\tilde{u}^{i}_L}^{2},M_{\tilde{u}^{j}_L}^{2},
M_{\tilde{u}^{j}_R}^{2}, M_{\tilde{u}^{i}_R}^{2}, |M_{g}|^{2}\right)\
\left(\delta^{u}_{ij}\right)_{RR} \left(\delta^{u}_{ji}\right)_{LL}\nonumber\\
&+&\frac{\left({\bf Y}_{u}\right)_{i i}}{(4\pi)^{2}}\ 
\left({\bf Y}_{d}^{\dagger}\right)_{i i} 
\left({\bf Y}_{d}\right)_{i i} |\mu|^{2}\ I_3\left(M_{\tilde{d}^{i}_R}^{2}, 
M_{\tilde{u}^{i}_L}^{2}, |\mu|^{2}\right)\nonumber\\
\Gamma^{u}_{i i} &=& \frac{2 \alpha_s}{3\pi}\ \left({\bf Y}_{u}\right)_{i i}
\mu^{\star} M_{g}^{\star}\ I_3\left(M_{\tilde{u}^{i}_L}^{2},
M_{\tilde{u}^{i}_R}^{2}, |M_{g}|^{2}\right)\nonumber\\
&+&\frac{2 \alpha_s}{3\pi} \sum_{j=1}^{3}
\left({\bf Y}_{u}\right)_{j j} \mu^{\star} M_{g}^{\star}\ M_{\widetilde{U}}^{4}
I_5\left(M_{\tilde{u}^{i}_L}^{2},M_{\tilde{u}^{j}_L}^{2},
M_{\tilde{u}^{j}_R}^{2}, M_{\tilde{u}^{i}_R}^{2}, |M_{g}|^{2}\right)\
\left(\delta^{u}_{ij}\right)_{RR} \left(\delta^{u}_{ji}\right)_{LL}\nonumber\\
&+&\frac{\left({\bf Y}_{u}\right)_{i i}}{(4\pi)^{2}}\ 
\left({\bf Y}_{d}^{A\, \dagger}\right)_{i i}
\left({\bf Y}_{d}\right)_{i i} \mu^{\star}\ I_3\left(M_{\tilde{d}^{j}_R}^{2},
M_{\tilde{d}^{i}_L}^{2}, |\mu|^{2}\right)
\end{eqnarray}
for the entries at the diagonal, and 
\begin{eqnarray}
\gamma^{u}_{i j} &=& \frac{2 \alpha_s}{3\pi}\
\left({\bf Y}_{u}^{A}\right)_{i j} M_{g}^{\star}\
I_3\left(M_{\tilde{u}^{j}_L}^{2},
M_{\tilde{u}^{i}_R}^{2}, |M_{g}|^{2}\right)\nonumber\\
&+&\frac{2 \alpha_s}{3\pi}\
\left({\bf Y}_{u}^{A}\right)_{i i} M_{g}^{\star} M_{\widetilde{U}}^{2}\
I_4\left(M_{\tilde{u}^{j}_L}^{2},
M_{\tilde{u}^{i}_L}^{2}, M_{\tilde{u}^{i}_R}^{2}, |M_{g}|^{2}\right)\: 
\left(\delta^{u}_{ij}\right)_{LL}\nonumber\\
&+&\frac{2 \alpha_s}{3\pi}\
\left({\bf Y}_{u}^{A}\right)_{j j} M_{g}^{\star} M_{\widetilde{U}}^{2}\
I_4\left(M_{\tilde{u}^{j}_L}^{2},
M_{\tilde{u}^{j}_R}^{2}, M_{\tilde{u}^{i}_R}^{2}, |M_{g}|^{2}\right)\:
\left(\delta^{u}_{ij}\right)_{RR}\nonumber\\
&+&\frac{\left({\bf Y}_{u}\right)_{i j}}{(4\pi)^{2}}\ 
\left({\bf Y}_{d}^{\dagger}\right)_{j j}
\left({\bf Y}_{d}\right)_{j j} |\mu|^{2}\ I_3\left(M_{\tilde{d}^{j}_R}^{2},
M_{\tilde{d}^{j}_L}^{2}, |\mu|^{2}\right)\nonumber\\
&+&\frac{\left({\bf Y}_{u}\right)_{i i}}{(4\pi)^{2}}\
\left({\bf Y}_{d}^{\dagger}\right)_{j j}
\left({\bf Y}_{d}\right)_{j j} |\mu|^{2} M_{\widetilde{D}}^2\
I_4\left(M_{\tilde{d}^{j}_R}^{2}, M_{\tilde{d}^{j}_L}^{2}, 
M_{\tilde{d}^{i}_L}^{2}, |\mu|^{2}\right)\: 
\left(\delta^{d}_{ij}\right)_{LL}\nonumber\\
&+&\frac{\left({\bf Y}_{u}\right)_{i i}}{(4\pi)^{2}}\
\left({\bf Y}_{d}^{\dagger}\right)_{i i}
\left({\bf Y}_{d}\right)_{j j} |\mu|^{2} M_{\widetilde{D}}^2\
I_4\left(M_{\tilde{d}^{j}_R}^{2}, M_{\tilde{d}^{i}_R}^{2},
M_{\tilde{d}^{i}_L}^{2}, |\mu|^{2}\right)\:
\left(\delta^{d}_{ij}\right)_{RR}\nonumber\\
\Gamma^{u}_{i j} &=& \frac{2 \alpha_s}{3\pi}\
\left({\bf Y}_{u}\right)_{i j} \mu^{\star} M_{g}^{\star}\
I_3\left(M_{\tilde{u}^{j}_L}^{2},
M_{\tilde{u}^{i}_R}^{2}, |M_{g}|^{2}\right)\nonumber\\
&+&\frac{2 \alpha_s}{3\pi}\
\left({\bf Y}_{u}\right)_{i i} \mu^{\star} M_{g}^{\star} M_{\widetilde{U}}^{2}\
I_4\left(M_{\tilde{u}^{j}_L}^{2},
M_{\tilde{u}^{i}_L}^{2}, M_{\tilde{u}^{i}_R}^{2}, |M_{g}|^{2}\right)\: 
\left(\delta^{u}_{ij}\right)_{LL}\nonumber\\
&+&\frac{2 \alpha_s}{3\pi}\
\left({\bf Y}_{u}\right)_{j j} \mu^{\star} M_{g}^{\star} M_{\widetilde{U}}^{2}\
I_4\left(M_{\tilde{u}^{j}_L}^{2},
M_{\tilde{u}^{j}_R}^{2}, M_{\tilde{u}^{i}_R}^{2}, |M_{g}|^{2}\right)\:
\left(\delta^{u}_{ij}\right)_{RR}\nonumber\\
&+&\frac{\left({\bf Y}_{u}\right)_{i j}}{(4\pi)^{2}}\ 
\left({\bf Y}_{d}^{A\, \dagger}\right)_{j j}
\left({\bf Y}_{d}\right)_{j j} \mu^{\star}\ I_3\left(M_{\tilde{d}^{j}_R}^{2},
M_{\tilde{d}^{i}_L}^{2}, |\mu|^{2}\right)\nonumber\\
&+&\frac{\left({\bf Y}_{u}\right)_{i i}}{(4\pi)^{2}}\
\left({\bf Y}_{d}^{A\, \dagger}\right)_{j j}
\left({\bf Y}_{d}\right)_{j j}  \mu^{\star} M_{\widetilde{D}}^2\
I_4\left(M_{\tilde{d}^{j}_R}^{2}, M_{\tilde{d}^{j}_L}^{2}, 
M_{\tilde{d}^{i}_L}^{2}, |\mu|^{2}\right)\: 
\left(\delta^{d}_{ij}\right)_{LL}\nonumber\\
&+&\frac{\left({\bf Y}_{u}\right)_{i i}}{(4\pi)^{2}}\
\left({\bf Y}_{d}^{A\, \dagger}\right)_{i i}
\left({\bf Y}_{d}\right)_{j j} \mu^{\star} M_{\widetilde{D}}^2\
I_4\left(M_{\tilde{d}^{j}_R}^{2}, M_{\tilde{d}^{i}_R}^{2},
M_{\tilde{d}^{i}_L}^{2}, |\mu|^{2}\right)\:
\left(\delta^{d}_{ij}\right)_{RR}
\end{eqnarray}
for the intergenerational ones. In these expressions
$M_{\widetilde{U},\widetilde{D}}$ stand for 
the average up and down squark masses, and
\begin{eqnarray}
\label{mis}
\left(\delta^{u,d}_{ij}\right)_{LL,RR}\equiv
\left(M_{U,D}^{2}\right)_{LL,RR}^{ij}/M_{\widetilde{U},\widetilde{D}}^{2}
\end{eqnarray}
are the mass insertions (MI) whose phases and sizes parametrize, respectively,
the CP and flavor violations from the intergenerational entries
of $\left(M_{U,D}^{2}\right)_{LL,RR}$.  Note that all 
entries of $\gamma^{u,d}$ and $\Gamma^{u,d}$ are computed
at one loop approximation, and SUSY flavor violation effects
are treated at single MI level everywhere except the diagonal 
entries which include dominant SUSY QCD contributions with two MIs
in addition to the leading zero MI diagrams. 
The radiative corrections depend on the loop functions $I_{3,4,5}$ where 
\begin{eqnarray}
\label{loopfunc}
I_n\left(m_1^2,m_2^2, \cdots ,m_n^2\right) &=& (-1)^{n+1}\Gamma(n-2) 
\int_{0}^{1} d x_1
\int_{0}^{1-x_1} d x_2 \cdots \int_{0}^{1-x_1-\cdots -x_{n-2}} d 
x_{n-1}\nonumber\\
&&\left(x_1 m_1^2 + x_2 m_2^2 + \cdots + (1-x_1-\cdots - x_{n-1}) m_n^2 
\right)^{-n}
\end{eqnarray}
which approach, respectively, to $1/2 m^2$, $-1/6 m^4$ and $1/12 m^6$
for $n=3,4$ and 5 when their arguments are equal.

An important aspect of the nonholomorphic Yukawa structures $\gamma^{u,d}$ and $\Gamma^{u,d}$
is that they depend only on the ratio of the soft masses not on their absolute scale. This
property guarantees that these radiative corrections remain nonvanishing even if
$M_{SUSY}\gg m_t$. The simplest case corresponds to an approximate universality
of the soft masses, $|\mu|\sim |M_{g}|\sim M_{\tilde{u}^{j}_{L,R}}\sim M_{\tilde{d}^{j}_{L,R}}\sim
|\left(A_{u,d}\right)_{i i}|\sim M_{\widetilde{U}}\sim M_{\widetilde{D}}\equiv M_{SUSY}$,
in which case $\gamma^{u,d}$ and $\Gamma^{u,d}$ depend only on the gauge and Yukawa couplings
in addition to CP and flavor violation textures from the soft masses. Altough such a universality
is not likely to occur at low energies even if it holds at the scale of local SUSY breaking, it
proves useful in illustrating the salient features of the Higgs interactions with quarks. Using
the limiting forms of the loop functions (\ref{loopfunc}) one obtains
\begin{eqnarray}
\label{limd}
\gamma^{d}_{i i} &\Longrightarrow& \left({\bf Y}_{d}\right)_{i i}\left[ \frac{\alpha_s}{3\pi}\ e^{i (\theta^{d}_{ii}-\theta_g)}
+\frac{1}{32\pi^{2}}\ \sum_{j=1}^{3}\ \left({\bf Y}_{u}^{\dagger}\right)_{i j}
\left({\bf Y}_{u}\right)_{j i}\right] \nonumber\\
&+& \frac{\alpha_s}{18\pi} \sum_{j=1}^{3} \left({\bf Y}_{d}\right)_{j j} 
\left(\delta^{d}_{ij}\right)_{RR} \left(\delta^{d}_{ji}\right)_{LL}\
e^{i (\theta^{d}_{jj}-\theta_g)}\nonumber\\
\Gamma^{d}_{i i} &\Longrightarrow& \left({\bf Y}_{d}\right)_{i i}\left[  
 \frac{\alpha_s}{3\pi}\ e^{-i (\theta_{\mu}+\theta_g)} +\frac{1}{32\pi^{2}}\ \sum_{j=1}^{3}\ 
\left({\bf Y}_{u}^{\dagger}\right)_{i j}\left({\bf Y}_{u}\right)_{j i}\ 
\left(\delta^{u}_{j i}\right)_{A}^{\star} 
e^{-i\theta_{\mu}}\right]\nonumber\\
&+&\frac{\alpha_s}{18\pi} \sum_{j=1}^{3}
\left({\bf Y}_{d}\right)_{j j} 
\left(\delta^{d}_{ij}\right)_{RR} \left(\delta^{d}_{ji}\right)_{LL}\
e^{-i(\theta_{\mu}+\theta_{g})}\nonumber\\
\gamma^{d}_{i j} &\Longrightarrow& \left({\bf Y}_{d}\right)_{i i}\ \left[ -\frac{\alpha_s}{9\pi}\ \left(\delta^{d}_{ij}\right)_{LL} e^{i 
(\theta^{d}_{ii}-\theta_g)}\right.\nonumber\\ &+&  
\left. \frac{1}{96\pi^2} \left\{ 3 \left({\bf Y}_{u}^{\dagger}\right)_{i j}
\left({\bf Y}_{u}\right)_{j j} + 3 \left({\bf Y}_{u}^{\dagger}\right)_{i i} \left({\bf Y}_{u}\right)_{i j}\right.\right.\nonumber\\
&-&\left.\left.\left({\bf Y}_{u}^{\dagger}\right)_{j j} \left({\bf Y}_{u}\right)_{j j}\ \left(\delta^{u}_{ij}\right)_{LL} 
- \left({\bf Y}_{u}^{\dagger}\right)_{i i} \left({\bf Y}_{u}\right)_{j j}\ 
\left(\delta^{u}_{ij}\right)_{RR} \right\}\right]\nonumber\\
&-&\left({\bf Y}_{d}\right)_{j j}\ \left[ \frac{\alpha_s}{9\pi} \left(\delta^{d}_{ij}\right)_{RR} e^{i (\theta^{d}_{jj}-\theta_g)}
\right]\nonumber\\  
\Gamma^{d}_{i j} &\Longrightarrow& \left({\bf Y}_{d}\right)_{i i}\ \left[ -\frac{\alpha_s}{9\pi}\ \left(\delta^{d}_{ij}\right)_{LL} e^{-i
(\theta_{\mu}+\theta_g)}\right.\nonumber\\ &+&
\left.\frac{1}{96\pi^2} \left\{ 3 \left({\bf Y}_{u}^{\dagger}\right)_{i j}
\left({\bf Y}_{u}\right)_{j j} \left(\delta^{u}_{j i}\right)_{A}^{\star} 
e^{-i\theta_{\mu}}
+ 3 \left({\bf Y}_{u}^{\dagger}\right)_{i i} \left({\bf Y}_{u}\right)_{i j} e^{-i(\theta^{u}_{ii}+\theta_{\mu})} \right.\right.\nonumber\\
&-&\left.\left.\left({\bf Y}_{u}^{\dagger}\right)_{j j} \left({\bf Y}_{u}\right)_{j j}\ \left(\delta^{u}_{ij}\right)_{LL} 
e^{-i(\theta^{u}_{jj}+\theta_{\mu})}
- \left({\bf Y}_{u}^{\dagger}\right)_{i i} \left({\bf Y}_{u}\right)_{j j}\ \left(\delta^{u}_{ij}\right)_{RR} 
e^{-i(\theta^{u}_{ii}+\theta_{\mu})}\right\}\right]\nonumber\\
&-&\left({\bf Y}_{d}\right)_{j j}\ \left[ \frac{\alpha_s}{9\pi} \left(\delta^{d}_{ij}\right)_{RR} e^{-i (\theta_{\mu}+\theta_g)}
\right]
\end{eqnarray}
for down quark sector, and 
\begin{eqnarray}
\label{limu}
\gamma^{u}_{i i} &\Longrightarrow& \left({\bf Y}_{u}\right)_{i i}\left[ \frac{\alpha_s}{3\pi}\ e^{i (\theta^{u}_{ii}-\theta_g)}
+\frac{1}{32\pi^{2}}\ \left({\bf Y}_{d}^{\dagger}\right)_{i i}
\left({\bf Y}_{d}\right)_{i i}\right] \nonumber\\
&+&\frac{\alpha_s}{18\pi} \sum_{j=1}^{3}
\left({\bf Y}_{u}\right)_{j j} 
\left(\delta^{u}_{ij}\right)_{RR} \left(\delta^{u}_{ji}\right)_{LL}
e^{i(\theta^{u}_{j j}-\theta_{g})}
\nonumber\\
\Gamma^{u}_{i i} &\Longrightarrow& \left({\bf Y}_{u}\right)_{i i}\left[
 \frac{\alpha_s}{3\pi}\ e^{-i (\theta_{\mu}+\theta_g)} +\frac{1}{32\pi^{2}}\ 
\left({\bf Y}_{d}^{\dagger}\right)_{i i}\left({\bf Y}_{d}\right)_{i i}
e^{-i(\theta^{d}_{ii}+\theta_{\mu})}\right]\nonumber\\
&+&\frac{\alpha_s}{18\pi} \sum_{j=1}^{3}
\left({\bf Y}_{u}\right)_{j j}
\left(\delta^{u}_{ij}\right)_{RR} \left(\delta^{u}_{ji}\right)_{LL}
e^{-i(\theta_{\mu}+\theta_{g})}\nonumber\\
\gamma^{u}_{i j} &\Longrightarrow& \left({\bf Y}_{u}\right)_{i j}\left[ \frac{\alpha_s}{3\pi}\ e^{i (\theta^{u}_{ij}-\theta_g)}
+\frac{1}{32 \pi^2} \left({\bf Y}_{d}^{\dagger}\right)_{j j}
\left({\bf Y}_{d}\right)_{j j}\right] \nonumber\\
&-&\left({\bf Y}_{u}\right)_{i i}\ \left[ \frac{\alpha_s}{9\pi}\ \left(\delta^{u}_{ij}\right)_{LL} e^{i
(\theta^{u}_{ii}-\theta_g)}\right.\nonumber\\ &+&
\left. \frac{1}{96\pi^2} \left\{ 
\left({\bf Y}_{d}^{\dagger}\right)_{j j} \left({\bf Y}_{d}\right)_{j j}\ \left(\delta^{d}_{ij}\right)_{LL}
+\left({\bf Y}_{d}^{\dagger}\right)_{i i} \left({\bf Y}_{d}\right)_{j j}\ \left(\delta^{d}_{ij}\right)_{RR} \right\}\right]\nonumber\\
&-&\left({\bf Y}_{u}\right)_{j j}\ \left[ \frac{\alpha_s}{9\pi} \left(\delta^{u}_{ij}\right)_{RR} e^{i (\theta^{u}_{jj}-\theta_g)}
\right]\nonumber\\
\Gamma^{u}_{i j} &\Longrightarrow& \left({\bf Y}_{u}\right)_{i j}\left[ \frac{\alpha_s}{3\pi}\ e^{-i (\theta_{\mu}+\theta_g)}
+\frac{1}{32 \pi^2} \left({\bf Y}_{d}^{\dagger}\right)_{j j}
\left({\bf Y}_{d}\right)_{j j} e^{-i(\theta^{d}_{jj}+\theta_{\mu})}\right] \nonumber\\
&-&\left({\bf Y}_{u}\right)_{i i}\ \left[ \frac{\alpha_s}{9\pi}\ \left(\delta^{u}_{ij}\right)_{LL} e^{-i
(\theta_{\mu}+\theta_g)}\right.\nonumber\\ &+&
\left. \frac{1}{96\pi^2} \left\{
\left({\bf Y}_{d}^{\dagger}\right)_{j j} \left({\bf Y}_{d}\right)_{j j}\ \left(\delta^{d}_{ij}\right)_{LL} e^{-i(\theta^{d}_{jj}+\theta_{\mu})}
+\left({\bf Y}_{d}^{\dagger}\right)_{i i} \left({\bf Y}_{d}\right)_{j j}\ \left(\delta^{d}_{ij}\right)_{RR} e^{-i(\theta^{d}_{ii}+\theta_{\mu})}
\right\}\right]\nonumber\\
&-&\left({\bf Y}_{u}\right)_{j j}\ \left[ \frac{\alpha_s}{9\pi} \left(\delta^{u}_{ij}\right)_{RR} e^{-i(\theta_{\mu}+\theta_g)}\right]
\end{eqnarray}
for up quark sector after introducing the CP--odd phases 
$\theta_{\mu}\equiv \mbox{Arg}[\mu]$, $\theta^{u,d}_{ii}\equiv 
\mbox{Arg}[\left(A_{u,d}\right)_{ii}]$, and  $\theta_{g}\equiv 
\mbox{Arg}[M_g]$. That ${\bf Y}_u$ is not diagonal causes all entries
of $A_{u}$ to contribute $\Gamma^{d}$, and this is parametrized via the
insertions $\left(\delta^{u}_{i j}\right)_{A} = \left(A_{u}\right)_{i 
j}/M_{SUSY}$, similar to (\ref{mis}). Note that in (\ref{limd}) and
(\ref{limu}) there is no explicit dependence on the soft masses except
for the fact that all Yukawa and gauge couplings are to be evaluated at
the scale $M_{SUSY}$.

The nonholomorphic Yukawa structures $\gamma^{u,d}$ and $\Gamma^{u,d}$
contribute to the quark masses as well as the Higgs boson couplings to
quarks after the electroweak breaking. When determining the vacuum
configuration and the physical Higgs bosons it is essential to include 
the radiative corrections to the Higgs potential from sparticle loops. 
In particular, the CP--odd phases contained in trilinear couplings
and the $\mu$ parameter generate sizeable scalar--pseudoscalar mixings
which prevent the Higgs bosons to have definite CP parities \cite{higgs}
(these results can be further refined using the most recent complete two
loop calculation \cite{martin}). Defining the Higgs VEVs as $v_{u,d}=\sqrt{2} \langle H^0_{u,d} \rangle$, with
$\tan\beta\equiv \langle H^0_{u} \rangle/\langle H^0_{d} \rangle$, the radiatively--corrected
Yukawa coupling matrices take the form 
\begin{eqnarray}
\label{corrYuk}
\Upsilon_d &=&   {\bf Y}_{d} - \gamma^d
+ \tan\beta\  \Gamma^d \:\:\:,\:\:\:\:
\Upsilon_u = e^{i\delta} \left[  {\bf Y}_{u} + \gamma^u
- \cot\beta\  \Gamma^u \right]
\end{eqnarray}
where $\delta$, the relative phase between the two doublets, is generated by the
SUSY CP phases via the radiative corrections \cite{shift}. These Yukawa matrices 
can be diagonalized via the rotations ${\bf d}_L\rightarrow V_L^d\ {\bf d}_L$, ${\bf u}_L\rightarrow V^{0
\dagger}\ V_L^u\ {\bf u}_L$, ${\bf d}_R\rightarrow
V_R^d\ {\bf d}_R$, and ${\bf u}_R\rightarrow V_R^u\ {\bf u}_R$. Then the
misalignment between the left--handed quarks in up and down sectors
generates the physical CKM matrix
\begin{eqnarray}
\label{ckm}
V=\left(V_L^u\right)^{\dagger}\ V^{0}\ V_L^d
\end{eqnarray}
which would be identical to $V^0$ in the absence of radiative corrections.
The defining relations for the unitary matrices $V^{u,d}_{L,R}$ are 
\begin{eqnarray}
\label{matrices}
&&\left(V_L^d\right)^{\dagger}\ \left(\Upsilon_d\right)^{\dagger}\ 
\Upsilon_d\  V_L^d = \overline{{\bf Y}_d}^{2} \:\:\:,\:\:\:
V\ \left(V_L^d\right)^{\dagger}\ \left(\Upsilon_u\right)^{\dagger}\ 
\Upsilon_u\ V_L^d\ V^{\dagger} = \overline{{\bf Y}_u}^{2}\nonumber\\
&&\left(V_R^d\right)^{\dagger}\ \Upsilon_d\ \left(\Upsilon_d\right)^{\dagger}\
V_R^d = \overline{{\bf Y}_d}^{2} \:\:\:,\:\:\:
\left(V_R^u\right)^{\dagger}\ \Upsilon_u\ \left(\Upsilon_u\right)^{\dagger}\
V_R^u = \overline{{\bf Y}_u}^{2}
\end{eqnarray}
where $\overline{{\bf Y}_d}=\mbox{diag.}\left\{\overline{h_d},
\overline{h_s}, \overline{h_b}\right\}$ and $\overline{{\bf Y}_u}=\mbox{diag.}\left\{\overline{h_u},    
\overline{h_c}, \overline{h_t}\right\}$ with $\overline{h_i}$  being the running (physical) 
Yukawa coupling of the $i$--th generation, $e.g.$ $\overline{h_s}=g_2\overline{m_s}/\sqrt{2} M_W \cos\beta$.
Note that in the above $\tan\beta$, $V$ as well as
$\overline{h_i}$ are all evaluated at the scale $M_{SUSY}$ via the 
RGE running of their experimental values at $M_Z$ using the two--Higgs doublet 
model as the effective theory below $M_{SUSY}$ \cite{pdg,evolve}. 
The mass--eigenstate quark fields above interact with the Higgs bosons via
\begin{eqnarray}
\label{interact}
-{\cal{L}}&=&\overline{{\bf d}_R}\ \overline{{\bf Y}_d}\ {\bf d}_L H_d^{0} +
\overline{{\bf d}_R}\ \left(V_R^d\right)^{\dagger}\ \Gamma^{d}\ V_L^d\ {\bf 
d}_L\ \left(H_{u}^{0 \star} - \tan\beta\ H_d^0\right)\nonumber\\
&+&\overline{{\bf u}_R}\ \overline{{\bf Y}_u}\ {\bf u}_L H_u^{0} -
\overline{{\bf u}_R}\ \left(V_R^u\right)^{\dagger}\ \Gamma^{u}\ V_L^d\ 
V^{\dagger}\ {\bf u}_L\ \left(H_{d}^{0 \star} - \cot\beta\ H_u^0\right)
\end{eqnarray}
where it is clear that the flavor structures of $\Gamma^{u,d}$ are crucial for Higgs bosons
to develop FCNC couplings. In particular, when $\Gamma^{d}\propto {\bf Y}_d$
and/or $\Gamma^{u}\propto {\bf Y}_u V^0$ there is no flavor--changing
couplings of the neutral Higgs bosons to down and/or up quarks.
The tree level CKM matrix $V^0$ is some unitary matrix which 
does not need to confront the experimental data unless the 
radiative effects contained in $\gamma^{u,d}$ and $\Gamma^{u,d}$
are vanishingly small. The allowed ranges for individual entries
of $V^0$ depend on the size of the SUSY flavor and CP violation effects since 
the physical CKM matrix (\ref{ckm}) must saturate at least the bounds from 
tree level FCNC processes \cite{pdg}. The mixing matrices $V^{u,d}_{L,R}$
can be computed via perturbation theory for small flavor mixings. On
the other hand, if the mixings are sizeable it is useful to employ
direct diagonalization by first transforming $\Upsilon_{u,d}$ into the
nearest--neighbour--interaction basis \cite{branco} and then solving
for Yukawa couplings and tree level CKM elements using the
techniques given in \cite{fritzsch}.

For determining the Higgs interactions with quarks (\ref{interact}) it is 
necessary to express the tree level parameters $\left( {\bf Y}_{u,d}, 
V^0\right)$ in terms of the physical ones $\left(\overline{{\bf Y}_{u,d}}, 
V\right)$ via (\ref{ckm}) and (\ref{matrices}). Since a given
entry of $\Upsilon_{u,d}$ depends on the Yukawa couplings of other
generations, a direct solution of (\ref{matrices}) will eventually
need a scanning of the parameter space by taking into 
account all the available constraints. However, for the purpose of 
illustrating the essential features of SUSY flavor and CP violation 
effects on Higgs--quark interactions it suffices to have  an approximate 
solution for Yukawa couplings, $i.e.$ one can neglect flavor 
mixings from $V^0$, and discard the SUSY electroweak
corrections all together which induces $\sim 20\%$ error in
estimating the bottom Yukawa. Furthermore, for compactness it 
is useful to use the limiting forms (\ref{limd},\ref{limu}) 
keeping in mind that size of the radiative corrections can be 
significantly altered if the universality assumption is relaxed.
Within these approximations, the Yukawa couplings admit the solutions
\begin{eqnarray}
\label{corryuk}
h_d &=& \frac{g_2\overline{m_d}}{\sqrt{2} M_W}\ \frac{\tan\beta}{1+ \epsilon \tan\beta}\  
\left[ 1 - \frac{\epsilon\ \tan\beta}{1+ \epsilon\ \tan\beta}\
\left\{ \frac{\overline{m_s}}{\overline{m_d}} \left(\delta^{d}_{12}\right)_{L R} +
\frac{\overline{m_b}}{\overline{m_d}} \left(\delta^{d}_{13}\right)_{L R} \right\} \right]\nonumber\\
h_s &=& \frac{g_2\overline{m_s}}{\sqrt{2} M_W}\ \frac{\tan\beta}{1+ \epsilon \tan\beta}\
\left[ 1 - \frac{\epsilon\ \tan\beta}{1+ \epsilon\ \tan\beta}\
\frac{\overline{m_b}}{\overline{m_s}} \left(\delta^{d}_{23}\right)_{L R}\right]\nonumber\\
h_b &=& \frac{g_2\overline{m_b}}{\sqrt{2} M_W}\ \frac{\tan\beta}{1+ \epsilon \tan\beta} 
\nonumber\\
h_u &=& \frac{g_2\overline{m_u}}{\sqrt{2} M_W}\
\left[ 1 - \epsilon_1\
\left\{ \frac{\overline{m_c}}{\overline{m_u}} \left(\delta^{u}_{12}\right)_{L R} +
\frac{\overline{m_t}}{\overline{m_u}} \left(\delta^{u}_{13}\right)_{L R} \right\} \right]\nonumber\\
h_c &=& \frac{g_2\overline{m_c}}{\sqrt{2} M_W}\ 
\left[ 1 - \epsilon_2\
\frac{\overline{m_t}}{\overline{m_c}} \left(\delta^{u}_{23}\right)_{L R}\right]\nonumber\\
h_t &=&  \frac{g_2\overline{m_t}}{\sqrt{2} M_W}
\end{eqnarray}
where the SUSY flavor violation contributions are separated from the ones which already exist 
in the minimal flavor violation (MFV) scheme \cite{higgsfcnc,higgsfcnc1}. These expressions for Yukawas have
been obtained by keeping only those terms not suppressed by $\tan\beta$ and linear in 
$\left(\delta^{d}_{ij}\right)_{L R}$. The new parameters in (\ref{corryuk}) are defined as 
$\epsilon=(\alpha_s/3\pi) e^{-i(\theta_{\mu}+\theta_g)}$, $\epsilon_{i}=(\alpha_s/3\pi) e^{i(\theta^{u}_{ii} 
-\theta_g)}$, and
\begin{eqnarray}
\left(\delta^{d}_{ij}\right)_{L R}=\frac{1}{6}\ \left(\delta^{d}_{ij}\right)_{R R} \left(\delta^{d}_{j 
i}\right)_{L 
L}\:\:\:,\:\:\:
\left(\delta^{u}_{ij}\right)_{L R}=\frac{1}{6}\  e^{i(\theta^{u}_{jj} -\theta^{u}_{ii})}\ \left(\delta^{u}_{ij}\right)_{R R} 
\left(\delta^{u}_{j i}\right)_{L L} 
\end{eqnarray}
which generate the effective LR transitions needed for 
correcting the diagonal Yukawa elements. In contrast to the MFV scheme, 
the Yukawa couplings acquire sizeable corrections from the those
of the heavier generations as suggested by (\ref{corryuk}). 
Indeed, the radiative corrections to
$h_d/\overline{h_d}$, $h_s/\overline{h_s}$, $h_u/\overline{h_u}$ and
$h_c/\overline{h_c}$ involve, respectively, the large factors
$\overline{m_b}/\overline{m_d}\sim (\tan\beta)_{max}^{2}$,
$\overline{m_b}/\overline{m_s}\sim (\tan\beta)_{max}$,
$\overline{m_t}/\overline{m_u} \sim (\tan\beta)_{max}^{3}$, and
$\overline{m_t}/\overline{m_c}\sim (\tan\beta)_{max}^{2}$
with $(\tan\beta)_{max}\simlt \overline{m_t}/\overline{m_b}$.
Unlike the light quarks, the top and bottom Yukawas remain stuck to
their MFV values to a good approximation. Therefore, the SUSY flavor
violation sources mainly influence the light sector whereby modifying
several processes they participate. The modifications in the Yukawa
couplings are important even at low $\tan\beta$ values. As an example,
consider $\left(\delta^{d}_{13}\right)_{L R}\sim 10^{-2}$
for which $h_d/h_d^{MFV} \simeq 0.02 (2.11), -2.3 (6.6),
-4.6 (17.7)$ for $\tan\beta=5, 20, 40$ at $\theta_{\mu}+\theta_g\leadsto 0 (\pi)$.
Note that the Yukawas are enhanced especially for $\theta_{\mu}+\theta_g\leadsto \pi$
which is the point preferred by Yukawa--unified models such as SO(10).

In general, as $\tan\beta\rightarrow (\tan\beta)_{max}$ the Yukawa
couplings of down and strange quarks become ampproximately degenerate with
the bottom Yukawa for $\left(\delta^{d}_{13, 23}\right)_{L R}\sim 0.1$
and $\theta_{\mu}+\theta_g\leadsto \pi$. There is no $\tan\beta$ enhancement
for up quark sector but still the large ratio $\overline{m_t}/\overline{m_u}$
sizeably folds $h_u$ compared to its SM value: $h_u \simeq 0.6\ e^{i(\theta^{u}_{11} - \theta_g)}\
\overline{h_c}$ with  $\left(\delta^{u}_{13} \right)_{L R}\sim 0.1$.
These spectacular enhancements in light quark Yukawas, 
though possible in a small corner of the parameter space, 
imply that the SUSY flavor violation effects can induce
strong modifications in light quark couplings to Higgs bosons --
an important aspect for both Higgs boson searches and FCNC processes
to be detailed below.

At this point one may wonder if the leptonic Yukawas can also be enhanced. By 
replacing the gluino--squark loops with bino--slepton loops, one finds that 
the radiative corrections are actually down by two orders of magnitude compared to the 
quark sector  even when $\tan\beta\sim (\tan\beta)_{max}$ and bino is nearly 
degenerate with sleptons. Moreover, if bino is the dark matter candidate these 
threshold  corrections are further suppressed. In summary, as follows
from (\ref{corryuk}), the SUSY flavor and CP violations modify the
hierarchy of the Yukawa couplings strongly even for small or moderate
$\tan\beta$ values. In fact, when $\tan\beta$ assumes its maximal value
and the MIs are ${\cal{O}}(1)$ one finds that ($i$) the down quark Yukawas acquires 
an approximate universality, ($ii$) the up quark Yukawa becomes degenerate
with the charm, and ($iii$) the Yukawa couplings of the third 
generation quarks, of the charm quark, and of all leptons remain 
stuck to their MFV values to a good approximation.

The couplings of Higgs bosons to quarks are fully determined by (\ref{interact}). The off--diagonal entries of $V^{d}_{R,L}$ in 
(\ref{matrices}) are approximately given by $-(1/3) \epsilon \tan\beta \left(\delta^{d}_{ij}\right)_{RR, LL}$. The corresponding 
entries of $V^{u}_{L,R}$ are down by a factor or $1/\tan\beta$. Clearly, for 
any regime of the parameter values, the texture 
of the tree level CKM matrix $V^{0}$ plays an important role in generating the physical CKM matrix $V$. 
As for an approximate analysis, one may take $V^{u,d}_{L,R}$ diagonal and neglect
scalar--pseudoscalar mixings in the Higgs sector \cite{higgs}. Note that errors
made in these approximations are sensitive to $\tan\beta$; therefore, they must be 
avoided in an accurate treatment of the problem. Modulo these 
approximations, assuming for simplicity a universal phase for 
$\left(A_d\right)_{ii}$ and $\left(A_u\right)_{ii}$ each, the couplings of the 
Higgs bosons to quarks take the form
\begin{eqnarray}
\label{higgsquark}
-{\cal{L}}&=& \frac{g_2 \overline{m_{d^i}}(M_{SUSY})}{2 M_W}\ \left[ \frac{h_d^i}{\overline{h_d^i}}\
\tan\beta\ C^{d}_a + \left(\frac{h_d^i}{\overline{h_d^i}} -1 \right) \left(e^{i(\theta^{d}_{ii}+\theta_{\mu})}\, C^{d}_{a} -
C^{u \star}_a \right)\right]\ \overline{d^i_R}\ d^i_L\ H_a \nonumber\\
&+& \frac{g_2 \overline{ m_{d^i}}(M_{SUSY})}{6 M_W}\ \epsilon \tan\beta
\left[ \frac{h_d^i}{\overline{h_d^i}}\ \left(\delta^{d}_{ij}\right)_{L L} +
\frac{h_d^j}{\overline{h_d^i}}\ \left(\delta^{d}_{ij}\right)_{R R} \right] 
\left( \tan\beta\ C^{d}_a - C^{u \star}_a\right)
\overline{d^i_R}\ d^j_L\ H_a \nonumber\\
&+& \frac{g_2 \overline{m_{u^i}}(M_{SUSY})}{2 M_W}\ \left[ C^u_a + e^{-i(\theta^{u}_{ii}+\theta_{\mu})}\,
\left( \frac{h_u^i}{\overline{h_u^i}}-1 \right)\, \left( C^{d \star}_a - 
\cot\beta\ C^u_a\right)\right] \overline{u^i_R}\ u^i_L\ H_a \nonumber\\
&+& \frac{g_2 \overline{m_{u^i}}(M_{SUSY})}{6 M_W}\ \epsilon
\left[ \frac{h_u^i}{\overline{h_u^i}}\ \left(\delta^{u}_{ij}\right)_{L L} +
\frac{h_u^j}{\overline{h_u^i}}\ \left(\delta^{u}_{ij}\right)_{R R} \right] \left( C^{d \star}_a - \cot\beta\ C^{u}_a\right)
\overline{u^i_R}\ u^j_L\ H_a
\end{eqnarray}
where $C^{d}_{a}\equiv \{ - \sin\alpha,\cos\alpha, i\sin\beta, -i\cos\beta\}$ and $C^{u}_{a}\equiv \{ \cos\alpha,
\sin\alpha, i\cos\beta, i\sin\beta\}$ in the basis $H_{a}\equiv \{h,H,A,G\}$. In deriving (\ref{higgsquark})
$\tan\beta$ is assumed to be large though not necessarily close to $(\tan\beta)_{max}$. That the MFV contributions
as well as ${\cal{O}}\left[\left(\tan\alpha_{ij}\right)^{2}\right]$ terms are absent in the flavor--violating couplings,
that the CKM matrix does not have a direct contribution, that the Higgs bosons assume well--defined CP parities, $\dots$  
are just the artefacts of the simplifying assumptions made above. These missing pieces can be incorporated into
the effective lagrangian by a more accurate analysis using the exact formulae derived before.

The effective lagrangian (\ref{higgsquark}) for Higgs--quark interactions
has a multitude of phenomenological implications covering hadronic, atomic 
as well as Higgs sytems. Quite generically, all the SUSY effects contained in 
(\ref{higgsquark}) scale as $1/M_A^2$ whereas the analagous effective 
lagrangian for gauge boson interactions with quarks as well as four--fermion 
operators generated by box diagrams 
do so as $1/M_{SUSY}^{2}$ \cite{fcnc1,fcnc1p}. Consequently, when the SUSY Higgs sector lies at
the weak scale the Higgs--quark interactions probe superpartners at all scales,
as dictated by ({\ref{limd}), ({\ref{limu}) and (\ref{higgsquark}), via their 
persistent effects on low energy observables. On the other hand, the non--Higgs
SUSY contributions can be important only when $M_{SUSY}$ lies around the
weak scale. When discussing the Higgs boson effects on various observables
it is convenient to separate the atomic and hadronic observables
from those in the Higgs system:

{\bf 1.} {\bf Implications for Hadronic and Atomic Systems}:\\
For such observables the Higgs boson effects filter
through Higgs mediation which may (FCNC observables) 
or may  not (atomic EDMs) require flavor violation.
Starting with the FCNC processes, one notes that $\Delta F = 2$ 
transitions proceed via double Higgs penguins. The dominant 
contributions come from the scalar operators of the form $\overline{h_R} l_L\
\overline{h_L} l_R$ where $(h,l)=(s,d), (c,u), (b,d)$ and $(b,s)$
for $K^0$--$\overline{K^0}$, $D^0$--$\overline{D^0}$, 
$B_d^0-\overline{B_d^0}$ and $B_s^0-\overline{B_s^0}$ mixings,
respectively. The Wilson coefficients of these operators follow
from the flavor--changing parts of (\ref{higgsquark}), and at large $\tan\beta$ read as 
\begin{eqnarray}
\label{c2lrK}
C_{2}^{LR}\left(K^0-\overline{K^0}\right)&=& -\frac{1}{9}\ |\epsilon|^{2}  \tan^{2}\beta \left(h_s 
\left(\delta^{d}_{21}\right)_{LL} + h_d \left(\delta^{d}_{21}\right)_{RR} \right)
\left(h_d^{\star} \left(\delta^{d}_{21}\right)_{LL} + h_s^{\star} \left(\delta^{d}_{21}\right)_{RR} 
\right)\nonumber\\
     &\times& \left( \frac{\sin^2(\alpha-\beta)}{M_H^2} + 
\frac{\cos^2(\alpha-\beta)}{M_h^2} + \frac{1}{M_A^2}\right)
\end{eqnarray}
with $C_{2}^{LR}\left(D^0-\overline{D^0}\right)=C_{2}^{LR}\left(K^0-\overline{K^0}\right)\left[\tan\beta\rightarrow 
1, h_d \rightarrow h_u, h_s \rightarrow h_c\right]$. The expressions for
$C_{2}^{LR}\left(B_{d,s}^0-\overline{B_{d,s}^0}\right)$ follow from 
(\ref{c2lrK}) with obvious replacements:
\begin{eqnarray}
\label{c2lrB}
C_{2}^{LR}\left(B_d^0-\overline{B_d^0}\right)&=& -\frac{1}{9}\ |\epsilon|^{2}  \tan^{2}\beta \left(h_b
\left(\delta^{d}_{31}\right)_{LL} + h_d \left(\delta^{d}_{31}\right)_{RR} \right)
\left(h_d^{\star} \left(\delta^{d}_{31}\right)_{LL} + h_b^{\star} \left(\delta^{d}_{31}\right)_{RR}
\right)\nonumber\\
 &\times& \left( \frac{\sin^2(\alpha-\beta)}{M_H^2} +  
\frac{\cos^2(\alpha-\beta)}{M_h^2} + \frac{1}{M_A^2}\right)
\end{eqnarray}
where it is clear from both (\ref{c2lrK}) and (\ref{c2lrB}) that the Higgs double 
penguins contribute to the CP violation in mixing -- a property not present in 
the MFV scheme \cite{higgsfcnc,higgsfcnc1}. For all four distinct meson systems
$C_{2}^{LR}$ is quadratic in the Yukawa coupling of the heaviest quark, and  
requires six MIs when the radiative corrections in (\ref{corryuk}) dominate. 
It is clear that strength of $C_{2}^{LR}$ depends on the absolute sizes
of MIs as well as relative phases between the LL and RR sector contributions. 

The Higgs exchange diagrams with a single flavor flip generate 
$\Delta F =1$ transitions of which $K_L\rightarrow \pi e^+e^-$, 
$B_d \rightarrow \phi K_s$,  $B_s\rightarrow \mu^+\mu^-$, 
$D\rightarrow \pi\pi\pi$, $B_d \rightarrow (\pi, K) \ell^+\ell^-$ 
form a few examples. For instance, at the matching scale the Higgs penguins generate
the scalar operator $\overline{u_R} c_L\ \overline{d_R} d_L$ with the coefficient 
\begin{eqnarray}
\label{c1lrK}  
C_{1}^{LR}\left(D\rightarrow \pi \pi \pi\right)&=& -\frac{1}{3} \epsilon\  h_d \
\left(h_u \left(\delta^{u}_{12}\right)_{LL} + h_c \left(\delta^{u}_{12}\right)_{RR} \right)\nonumber\\
   &\times&  \left( \frac{\sin^2(\alpha-\beta)}{M_H^2} +
\frac{\cos^2(\alpha-\beta)}{M_h^2} + \frac{1}{M_A^2}\right)
\end{eqnarray}
which is responsible for $D$ meson decays into three pions. Similarly, semileptonic
operator $\overline{s_R} b_L \overline{\ell_{R}} \ell_L$, generated by Higgs 
exchange, contributes to pure leptonic decay of $B_{d}$ meson via
\begin{eqnarray}
\label{c1lrB}   
C_{1}^{LR}\left(B_{d}\rightarrow \overline{\mu}\mu\right)&=& -\frac{1}{3} \epsilon \tan\beta\  h_{\mu} \
\left(h_d \left(\delta^{d}_{13}\right)_{LL} + h_b \left(\delta^{d}_{13}\right)_{RR} \right)\nonumber\\
   &\times&  \left( \frac{\sin^2(\alpha-\beta)}{M_H^2} +
\frac{\cos^2(\alpha-\beta)}{M_h^2} + \frac{1}{M_A^2}\right)
\end{eqnarray}
with a similar expression for $B_s$ mode. It is clear that the strength of $C_1^{LR}$ 
is directly correlated with the associated $C_2^{LR}$ 
coefficient \cite{higgsfcnc,higgsfcnc1,higgsfcnc2}.

The EDMs of heavy atoms are sensitive to CP--violating 
semileptonic four--fermion operators \cite{lp,pilaftsisx,barr} in addition to
the electron EDM contribution. Especially operators of the form 
$\overline{q} q\ \overline{e} i \gamma_5 e$ couple the spin of the
electron cloud to the nuclear density, and the resulting contribution
to the EDM of the atom grows with its atomic number. For example,
the EDM of $^{205}$Tl is given by $d_{Tl} = -585\ d_e - 8.5\times 10^{-17}\ {C}_{S}\ {\rm e-cm}$,
where $d_e$ is the electron EDM and 
\begin{eqnarray}
\label{CS}
{C}_{S} = 5.5\times 10^{-10}\ \left(\frac{100\ {\rm GeV}}{M_A}\right)^{2}\ \mbox{Im}\left[
          (1-0.25 \kappa) h_b^{\star} h_e + 3.3 \kappa h_s^{\star} h_e + 0.5 h_b^{\star} h_e\right] 
\end{eqnarray}
with $\kappa \sim 1$ parametrizing the uncertainity in $\langle N | m_s \overline{s} s | N\rangle$. 

The hadronic and atomic system observables examplified by (\ref{c2lrK})--(\ref{CS})
are partly under experimental investigation and are partly constrained by the existing
data. In general, the bounds on these Higgs--exchange amplitudes depend on the size  
of non--Higgs contributions to a given observable. Conversely, the
existing bounds on various SUSY parameters \cite{fcnc1p,fcnclast}
derived by considering only the non--Higgs effects can be
significantly modified once the Higgs mediation effects are taken
into account. This has already been shown to happen
for the atomic EDMs \cite{pilaftsisx}: the Higgs--exchange
amplitude largely cancels with the two--loop electron EDM
contribution in certain regions of the parameter space. 
Therefore, it is after a combined analysis of the Higgs and non--Higgs
contributions that one can arrive at conclusions about the size and
phase contents of various MIs. Indeed, the present bounds on various
flavor violation sources \cite{fcnc1p,fcnclast}, following from
meson mixings by taking into account only the gluino box contributions,
can be relaxed or strengthened depending on the parameter space.

Among various FCNC observables, the pure leptonic decays of $B$ mesons 
put stringent constraints on Higgs--mediated FCNC since 
the SM predictions for $\mbox{BR}(B_{d,s}\rightarrow \overline{\mu}\mu)
\sim (1.5, 35) \times 10^{-10}$ which are roughly three orders
of mangitude below the current experimental bounds $(6.8, 26) \times 10^{-7}$ 
\cite{bmumu} whereas the Higgs--exchange contributions well 
exceed the bounds even in minimal flavor violation scheme for 
$\tan\beta\simgt 50$ \cite{higgsfcnc}. Due to the smallness of
the SM background the Higgs effects on these decays are important
irrespective of if $M_{SUSY}$ is close to or far above the weak scale. In 
the present framework, to agree with the bounds the Wilson coefficient
(\ref{c1lrB}) must be suppressed in other words the quantity
${h_d^i}\ \left(\delta^{d}_{ij}\right)_{L L} +{h_d^j}\ 
\left(\delta^{d}_{ij}\right)_{R R}$  (with $i=3$, $j=1,2$
and vice versa) must be sufficiently small depending on 
$\tan\beta$ and $M_A$. This constraint is important since
if it forces $\left(\delta^{d}_{13, 23}\right)_{L L, R R}$ 
to take small values the light quark Yukawas cannot assume
sizeable enhancements noted before.
Since (\ref{higgsquark}) is far from being precise enough (neglect
of flavor violation from $V^0$ and $V^{d}_{L,R}$ as well as the
SUSY electroweak corrections) to perform a scanning of the 
parameter space, it is useful check if (\ref{c1lrB}) can be
suppressed in parameter regions with low $M_A$, large $\tan\beta$
and ${\cal{O}}(1)$ MIs. This indeed happens. To see this one
incorporates terms higher order in MIs into the Yukawa
couplings listed in (\ref{corryuk}). For instance, the down quark
Yukawa takes the form
\begin{eqnarray}
h_d = h_d^{MFV}\ \frac{1 - a^2 \left(\delta^{d}_{23}\right)_{LR}\left(\delta^{d}_{32}\right)_{LR}
- a A_{12}\ \frac{\overline{m_s}}{\overline{m_d}} - a A_{13}\ \frac{\overline{m_b}}{\overline{m_d}}}
{1 - a^2 A_2 - a^3 A_3}
\end{eqnarray}
where $a= \epsilon \tan\beta/(1+ \epsilon \tan\beta)$, $A_{12}=\left[\left(\delta^{d}_{12}\right)_{LR} - a 
\left(\delta^{d}_{13}\right)_{LR} \left(\delta^{d}_{32}\right)_{LR}\right]$, 
$A_{13}=A_{12}(2\leftrightarrow 3)$, $A_2=\left|\left(\delta^{d}_{12}\right)_{LR}\right|^{2}+   
\left|\left(\delta^{d}_{13}\right)_{LR}\right|^{2}+\left|\left(\delta^{d}_{23}\right)_{LR}\right|^{2}$,
and $A_3=\left(\delta^{d}_{12}\right)_{LR} \left(\delta^{d}_{23}\right)_{LR}
\left(\delta^{d}_{31}\right)_{LR} + \mbox{h.c}$. Using these improved expressions
for Yukawas in (\ref{higgsquark}), one finds that the flavor--changing Higgs 
vertices $b s H^{a}$ and $b d H^{a}$ become
vanishingly small for $\tan\beta\simeq 60$ when all MIs are ${\cal{O}}(1)$, 
for $\tan\beta\simeq 65$ when $\left(\delta^{d}_{12}\right)_{LL,RR} \simeq 0$, 
and finally for $\tan\beta\simeq 68$ when $\left(\delta^{d}_{12}\right)_{LL} 
\simeq - \left(\delta^{d}_{12}\right)_{RR}$ provided that 
$\phi_{\mu} + \phi_{g} \leadsto \pi$ in all three cases. The existence of
such a parameter domain is important in that it allows one to overcome 
constraints from $B_{d,s}\rightarrow \overline{\mu} \mu$ without suppressing
the MIs $\left(\delta^{d}_{13,23}\right)_{LL,RR}$ which are crucial for
enhancing the light quark Yukawas. Of course, saturating the bounds implies 
by no means vanishing of $b s H^{a}$ and $b d H^{a}$ vertices instead 
what is nedeed is to suppress such flavor--changing entries without enforcing
the MIs to unobservably small values. If $M_{SUSY} \gg m_t$ vanishing
of such vertices reduces $B_{d,s} \rightarrow \overline{\ell}\ell$ and
$B_{d,s} \rightarrow (K, \pi) \ell^+\ell^-$ decays as well as 
$B^0_{d,s}$--$\overline{B^0_{d,s}}$ mixings to their SM predictions
with ${\cal{O}}(1)$ flavor mixings between the third and first generations
of quarks. If $M_{SUSY} \sim m_t$, however, for flavor mixings to be still
sizeable the Higgs exchange contributions to $B^0_{d,s}$--$\overline{B^0_{d,s}}$
should balance the  gluino boxes \cite{fcnc1p,fcnclast} on top of suppressing $B_{d,s}
\rightarrow \overline{\mu}\mu$ below the bounds. This can be
decided only after a global analysis of all the existing FCNC data.

For flavor transitions between the first two generations, the relevant
observables are $K^0$--$\overline{K^0}$  and $D^0$--$\overline{D^0}$ mixings
as well as the rare $K$ and $D$ decays. The Higgs--exchange 
amplitudes (\ref{c2lrK}) and (\ref{c1lrK}) can be suppressed
either by balancing them with the gluino boxes or by tuning
the LL and RR pieces in (\ref{higgsquark}) depending, respectively,
on whether $M_{SUSY}\sim m_t$ or  $M_{SUSY}\gg m_t$. One notices
that, in the latter case, the equality $\left(\delta^{u,d}_{12}\right)_{LL}\simeq 
- \left(\delta^{u,d}_{12}\right)_{RR}$ automatically suppresses
the FCNC in $K$ and $D$ systems without falsifying the aforementioned
enhancements in light quark Yukawas which rest mainly on 
$\left(\delta^{u,d}_{13,23}\right)_{LR}$.

The atomic EDMs, in particular, $^{205}Tl$ EDM, can be 
suppressed by balancing the contribution of (\ref{CS}) 
with $d_e$ (especially its two--loop part) \cite{pilaftsisx}
if $M_{SUSY}$ is close to the weak scale. In the opposite
limit, $M_{SUSY} \gg m_t$, (\ref{CS}) is the only 
contribution and its suppression is almost automatic
in parameter regions where the Higgs--mediated FCNC
in $B$ system are suppressed.

{\bf 2.} {\bf Implications for Higgs Boson Searches}:\\
The sizes and phases of various MIs and other SUSY parameters
that have survived the bounds from EDMs and FCNC observables 
can open new channels, strengthen or weaken the existing ones for Higgs
boson production and decays. The relevant interactions can be
read off from (\ref{higgsquark}) for each quark and Higgs flavor.
Concerning the collider searches for a light fundamental scalar,
the main object is the lightest Higgs boson which possesses both
flavor--changing and flavor--conserving couplings to quarks
\begin{eqnarray}
\label{int}
\mbox{Re}\left[g^{d}_{ii}\right] h \overline{d^i} d^i + 
\mbox{Re}\left[g^{u}_{ii}\right] h \overline{u^i} u^i + 
\frac{1}{2}\left\{ \left(g^{d}_{ij}+g^{d \star}_{ji}\right) h \overline{d^i} 
d^j + \left(g^{u}_{ij}+g^{u \star}_{ji}\right) h \overline{u^i} u^j + 
\mbox{h.c.}\right\}
\end{eqnarray}
where $i\neq j$, and various couplings read as 
\begin{eqnarray}
g^{d}_{ii} &=& - \left(h_d^i\right)_{SM}\
\frac{\sin\alpha}{\cos\beta}\ \left[1 + \left(\frac{h_d^i}{\overline{h_d^i}}
-1 \right) \left( 1 + \frac{1}{\tan\alpha \tan\beta}\right) \right]\nonumber\\
g^{d}_{ij} &=& - \left(h_d^i\right)_{SM}\
\frac{\sin\alpha}{\cos\beta}\ \frac{\epsilon}{3}\ 
\left[\frac{h_d^i}{\overline{h_d^i}} \left(\delta^{d}_{ij}\right)_{L L}
+ \frac{h_d^j}{\overline{h_d^i}} \left(\delta^{d}_{ij}\right)_{R R} \right]
\left(\tan\beta + \cot\alpha\right)\nonumber\\
g^{u}_{ii} &=& \left(h_u^i\right)_{SM}\
\frac{\cos\alpha}{\sin\beta}\ \left[1 + \left(1-\frac{h_u^i}{\overline{h_u^i}}
\right)\ e^{i(\theta^{u}_{ii}+\theta_{\mu})}\ \left(\cot\beta + 
\tan\alpha\right) \right]\nonumber\\
g^{u}_{ij} &=& \left(h_u^i\right)_{SM}\
\frac{\cos\alpha}{\sin\beta}\ \frac{\epsilon}{3}\ \left[
\frac{h_u^i}{\overline{h_u^i}} \left(\delta^{u}_{ij}\right)_{L L}
+ \frac{h_u^j}{\overline{h_u^i}} \left(\delta^{u}_{ij}\right)_{R R} \right]
\left(\cot\beta + \tan\alpha\right)
\end{eqnarray}
where the Yukawa couplings are given in (\ref{corryuk}). Clearly,
all flavor--diagonal couplings reduce to those in the SM, and 
all flavor--changing couplings vanish when the Higgs sector 
enters the decoupling regime, $M_{A}\gg M_Z$ \cite{hff}. In this
limit the lightest Higgs becomes the standard Higgs boson, and
all the aforomentioned Higgs--mediated FCNC amplitudes, which are generated by the heavier 
Higgs bosons, are suppressed as $1/M_A^2$. Therefore, only with a 
light Higgs sector, $i.e.$ $M_A\simgt M_Z$ or equivalently
$|\cot\alpha|\ll \tan\beta$, that there exist observable SUSY 
effects in the decay channels of $h$.

The Higgs bosons possess both flavor--changing and
flavor--conserving decay and production modes. For example,
the lightest Higgs decays into down and strange quarks in both
flavor--changing 
\begin{eqnarray}
\label{hds}
\frac{\Gamma(h\rightarrow \overline{s} d + \overline{d} s)}{\Gamma(h\rightarrow \overline{b} b)}
\simeq \left| \frac{\epsilon h_s + \epsilon^{\star} h_d^{\star}}{3 h_b}\ \left(\delta^{d}_{21}\right)_{L 
L} + \frac{\epsilon h_d + \epsilon^{\star} h_s^{\star}}{3 h_b}\ 
\left(\delta^{d}_{21}\right)_{RR}\right|^{2}\
\tan^2\beta
\end{eqnarray}
and flavor--conserving fashion
\begin{eqnarray}
\label{hdd}
\frac{\Gamma(h\rightarrow \overline{d} d)}{\Gamma(h\rightarrow \overline{b} b)}\simeq
\left(\mbox{Re}\left[\frac{h_d}{{h_b}}
\right]\right)^{2}
\end{eqnarray}
where the differences in phase spaces are neglected, and it is assumed that 
the Higgs sector is light $i.e.$ $|\cot\alpha|\ll \tan\beta$. Whether the 
ratios above achieve any observable significance depends on the sizes and
phases of $h_{d,s}$ as well as $\left(\delta^{d}_{21}\right)_{L L, R R}$,
which eventually need a global analysis of  all the relevant FCNC data.

It is useful to start analyzing (\ref{hds}) and (\ref{hdd}) in the limit
of enhanced Yukawas: $h_d\sim h_s \sim h_b$ and  $h_u \sim h_c$. In the
limit of heavy superpartners, $M_{SUSY} \gg m_t$, bounds on $B$, $D$ and
$K$ system FCNC can be satisfied with ${\cal{O}}(1)$ MIs, as discussed
above. In this case, flavor--changing Higgs interactions can be sizeable
only in the FCNC top decays \cite{tch} with $\Gamma(t \rightarrow c h) \simeq
\Gamma(t \rightarrow u h)$, whose likelihood depends on future observations
at Tevatron and LHC. Although all flavor--changing Higgs decay channels
are shut--off by the FCNC data, decays into $\overline{q} q$ final states
are maximally enhanced. Indeed, (\ref{hdd}) implies that $\Gamma(h\rightarrow
\overline{d}d) \simeq \Gamma(h\rightarrow \overline{s}s)  \simeq 
\Gamma(h\rightarrow \overline{b}b)$ and $\Gamma(h\rightarrow \overline{u}u)
\simeq \Gamma(h\rightarrow \overline{c}c)$. Therefore, $h\rightarrow \overline{b} b$
is no longer the dominant decay channel as expected in SM, instead all
channels are equally possible.  In fact, $\mbox{BR}(h\rightarrow \overline{b} b)$
is typically $\sim 30 \%$ which well below the SM expectation.
One recalls that, in the parameter domains which lead to degenerate 
Yukawas $\Gamma(h\rightarrow \overline{b}b)$ is typically an 
order of magnitude larger than that in the SM \cite{hollik}, and thus,
the main signature of enhanced down Yukawas is not the suppression
of $\overline{b}b$ production rate istead it is the drop in 
$\mbox{BR}(h\rightarrow \overline{b} b)$ due to the strengthening of
the other channels.  Note that, even the existing LEP data can accomodate a light Higgs boson with mass 
$\simlt 100\ {\rm GeV}$ provided that the Yukawa couplings are comparable in size \cite{lighthiggs} in
constrast to SM--like hierarchical couplings \cite{lephiggs}. 

When $M_{SUSY}$ is close to the weak scale, the FCNC observables receive 
contributions from not only the Higgs mediation but also from SUSY box and 
penguin diagrams. In this case, the allowed sizes of the MIs depend on
if Higgs and non--Higgs contrbutions sufficiently cancel. Note that 
even if this happens one still needs to suppress the pure leptonic
decay modes $B_{s,d}\rightarrow \ell^+\ell^-$ by tuning the SUSY 
parameters. This necessarily suppresses the corresponding flavor--changing
Higgs decays $h\rightarrow \overline{b} (s,d) + \overline{(s,d)} b$ \cite{herrero}.
In case all the MIs survive FCNC constraints without excessive 
suppression, then (\ref{hds}) and (\ref{hdd}) can both
be sizeable and therefore $\mbox{BR}(h\rightarrow \overline{b} b)$
can fall to $15$--$20 \% $ level as an optimistic  estimate.

Having discussed the implications of FCNC data for Higgs decays and
production for enhanced light quark Yukawas, it is useful to 
discuss (\ref{hds}) and (\ref{hdd})
in a different parameter domain, $i.e.$ suppose, though not realistic at all,
that the Higgs--exchange contributions to FCNC data are negligibe
so that the MIs remain stuck to their bounds obtained  
via non--Higgs amplitudes:
$\left(\delta^{d}_{12}\right)_{L L} \simeq \left(\delta^{d}_{12}\right)_{RR}
\simeq 8.0 \times 10^{-2}$, $\left(\delta^{d}_{13}\right)_{LL} \simeq 
\left(\delta^{d}_{13}\right)_{RR}
\simeq \left(\delta^{u}_{12}\right)_{LL} \simeq 
\left(\delta^{u}_{12}\right)_{RR} \simeq \left(\delta^{u}_{13} 
\right)_{LL} \simeq 2.0 \times 10^{-1}$,
and $\left(\delta^{d}_{23} \right)_{LL}\simeq \left(\delta^{d}_{23} \right)_{RR} \simeq
\left(\delta^{u}_{23} \right)_{LL} \simeq  \left(\delta^{u}_{23} \right)_{RR} \simeq 
\left(\delta^{u}_{13}\right)_{RR} \simeq 1$  as follows from the analyses of  \cite{fcnc1p,fcnclast} for 
$M_{SUSY}=1\ {\rm TeV}$. For definiteness, take $\theta_{\mu}+\theta_g  \leadsto \pi$, and  
consider $\tan\beta=20$ and $\tan\beta=\overline{m_t}/\overline{m_b}\simeq 60$ as two sample points
for illustrating low and high $\tan\beta$ behaviours. Then the down and strange quark Yukawas
are enhanced as $h_d=(2.4\times 10^{-3}, 0.04)\ h_b$ and $h_s=(0.06, 0.91)\ h_b$  for 
$\tan\beta=(20,60)$. Consequently, (\ref{hdd}) gives  $\Gamma(h \rightarrow \overline{d} d)=(6\times 
10^{-4}\ \%, 0.14\ \%)\ \Gamma(h \rightarrow \overline{b} b)$, $\Gamma(h \rightarrow \overline{s} 
s)=(0.35\ \%,  83\ \%)\ \Gamma(h \rightarrow \overline{b} b)$. Similarly, from (\ref{hds}) it follows 
that $\Gamma(h\rightarrow \overline{s} d + \overline{d} s)
=(7.2 \times 10^{-5}\ \%, 0.1\ \%)\ \Gamma(h \rightarrow \overline{b} b)$, $\Gamma(h\rightarrow 
\overline{b} d + \overline{d} b) =(0.12\ \%, 1.1\ \%)\ \Gamma(h \rightarrow \overline{b} 
b)$, and $\Gamma(h\rightarrow \overline{b} s + \overline{s} b)
=(16\ \%, 96\ \%)\ \Gamma(h \rightarrow \overline{b} b)$. It is clear that enhancements in
Yukawas depend crucially on the allowed sizes of the MIs as well as $\tan\beta$: at 
low $\tan\beta$ $h\rightarrow \overline{b} b$ is the dominant decay channel with a large
branching fraction as in the SM. On the other hand, as $\tan\beta$ grows to its maximal
value, $h\rightarrow \overline{s} s$ as well as $h\rightarrow \overline{b} s + \overline{s} b$
become as strong as  $h\rightarrow \overline{b} b$ since $\left(\delta^{d}_{23} \right)_{LL,RR}$
are ${\cal{O}}(1)$. In particular, one notes that $h\rightarrow \overline{b} b$ branching fraction
falls to $\sim 30 \%$ -- a completely non--SM signal testable in present \cite{lighthiggs,lephiggs}
as well as future colliders. Note that, though $ \Gamma(h \rightarrow \overline{d} d, \overline{s} s)
\ll  \Gamma(h \rightarrow \overline{b} b)$ for $\tan\beta=20$ they are still an order of magnitude 
larger than the SM prediction. The above analysis can also be repeated for the up quark sector. For
instance, taking $\theta^{u}_{ii}-\theta_{g} \leadsto \pi$ one finds $h_u \simeq 2.3\times 10^{-2}\ 
\overline{h_c}$ and $h_c \simeq 1.7 \overline{h_c}$ so that the most important enhancement occurs
for $h \rightarrow \overline{c} c$ decay whose rate is roughly four times larger than the SM 
expectation. Similarly, with $\left(\delta^{u}_{23} \right)_{LL} \simeq  \left(\delta^{u}_{23} 
\right)_{RR} \simeq \left(\delta^{u}_{13}\right)_{RR} \simeq 1$ one expects $\Gamma(t \rightarrow
c h) \simeq 4 \Gamma(t \rightarrow u h)$ whose absolute size depends on  how large $\tan\alpha$
is as follows from (\ref{int}). In spite of all these estimates for Higgs decay and production
rates, one keeps in mind that the MIs used above have been determined \cite{fcnc1p,fcnclast}
by discarding the Higgs--exchange amplitudes which grow with $\tan\beta$. Therefore, 
it is after a complete analysis of various FCNC and EDM observables by including all 
Higgs as well as non--Higgs contributions that one can achieve an accurate determination
of Higgs boson decay and production rates. 

Here it must be emphasized that the discussions above 
neglect the radiative corrections in the Higgs sector which are, however, important
and must be taken into account in an accurate analysis of the aforementioned 
observables because ($i$) at large $\tan\beta$ the radiative corrections can suppress
the $H^0_u$--$H^0_d$ mixing in the Higgs mass--squared matrix
so that the lightest Higgs can become effectively blind to all
down type fermions \cite{wells,higgs}, and ($ii$) the SUSY CP violation in the
Higgs sector modifies Higgs couplings to quarks and vector bosons
thereby altering the Higgs production and decay processes \cite{higgs,higgsprod}.

The discussions presented in the text show that the SUSY CP and
flavor violation effects can have important implications for
atomic EDMs, rare processes as well as the collider searches for
Higgs bosons. Several effects; sizeable modifications in light
quark Yukawas, filtering of SUSY CP violation into the meson
mixings, enhancements and certain regularities in Higgs boson decay
and production rates $\dots$ all induce
observable effects at meson factories and colliders.

\vspace{0.3cm}

The author greatfully acknowledges helpful conversations with Misha Voloshin
and Tonnis ter Veldhuis, and several discussions on atomic EDMs and FCNC phenomena in MFV scheme 
with Keith Olive and Maxim Pospelov. He also thanks Oleg Lebedev for
useful e-mail exchanges on MFV atomic EDMs, Howard Haber for
his comments on the decoupling regime in the Higgs sector, Gordon Kane
for discussions on Higgs decays, and  Maria Herrero for bringing
Ref.\cite{herrero} to his attention. This work is supported in part by the DOE grant 
DE-FG02-94ER40823.

\end{document}